\documentclass[12pt]{iopart}
\usepackage{iopams} 
\usepackage[latin1]{inputenc}
\usepackage{graphicx}
\usepackage{url}

\begin{document} 
 
\title[Goos-H\"anchen shift near the excitation
of backward SPPs]{Critical coupling layer thickness for positive or negative Goos-H\"anchen 
shifts near the excitation of backward surface polaritons in Otto-ATR systems}
\author{Mariana A. Zeller$^{(1)}$, Mauro Cuevas$^{(2)}$ and Ricardo A. Depine$^{(1)}$}
%\author{Mariana A. Zeller$^{(1)}$, Mauro Cuevas$^{(2)}$ and Ricardo A. Depine$^{(1)}$
\address{(1) Grupo de Electromagnetismo Aplicado, Departamento de F\'{\i}sica, FCEN, Universidad de Buenos Aires and IFIBA, Consejo Nacional de Investigaciones Cient\'{\i}ficas y T\'{e}cnicas (CONICET), Ciudad Universitaria, Pabell\'{o}n I, C1428EHA, Buenos Aires, Argentina. \\
(2) Facultad de Ingenier\'{\i}a, Universidad de Belgrano, Buenos Aires, Argentina}
\ead{rdep@df.uba.ar}

\begin{abstract} 
A theoretical analysis of the lateral displacement (Goos-H\"anchen shift) of spatially limited beams reflected from Attenuated Total Reflection (ATR) devices in the Otto configuration is presented when 
backward surface plasmon polaritons are excited at the interface between a positive refractive index slab and a semi--infinite metamaterial with negative refractive index. 
First, the stationary phase approximation and a phenomenological model based on the properties of the complex poles and zeroes of the reflection coefficient are used to demonstrate that: i) the excitation of backward surface waves can lead to both negative and positive (and not exclusively negative) Goos-H\"anchen shifts, and ii) the sign of the shift depends on whether the value of the coupling layer thickness is higher or lower than a critical value characteristic of the ATR structure. 
Second, these  findings are verified through rigorous calculations of the spatial structure of the reflected beam. 
For incident beams with a Gaussian profile, the lateral shift calculated as the first moment of the field distribution of the reflected beam agrees quite well with the predictions of approximate analysis. Near the resonant excitation of the backward surface plasmon polariton, large (negative or positive) Goos-H\"anchen shifts are obtained, along with a splitting of the reflected beam. 
\end{abstract} 

\section{Introduction} 

The Goos-H\"anchen (GH) lateral shift is an optical phenomenon in which a spatially limited beam reflected at an interface near conditions of total internal reflection is displaced from the incident beam position. Although conjectured by Isaac Newton in the 18th century \cite{Newton1}, experimental evidence was only provided in 1943 by Hilda H\"anchen (a doctoral student at the time) and her thesis advisor Fritz Goos \cite{Hanchentesis,Goos-Hanchen1}. 
The effect was theoretically explained by Artmann \cite{Artmann}, who noted that each plane wave component in the expansion of the incident beam undergoes a slightly different phase change after total internal reflection, so the sum of all the reflected components produces a lateral displacement of the reflected beam from the geometrical optics prediction. When the stationary phase approximation is used to calculate the sum of all the reflected components, the lateral beam shift is given by 
\begin{equation}
\Delta = - \frac{d\varphi}{d\alpha},
\label{stationary}
\end{equation}
where $\alpha$ is the tangential component of the wave vector in the medium of incidence and $\varphi$ is the phase difference between the reflected and incident waves.
% agregado para referee 2
Incidentally, we note that Artmann's argument shows that a prerequisite for the existence of a Goos-H\"anchen shift is not the excitation of surface or evanescent waves but rather the phases of the different plane wave components in the expansion of the incident beam being shifted differently upon reflection. 

The GH lateral shift and other nonspecular phenomena \cite{Nasalski2}, such as beam profile deformation, focal displacement, spatial waist modification and angular deviation, have been theoretically investigated in diverse configurations which  include multilayers \cite{Tamir1} and periodic structures \cite{Zhang}. 
For beams reflected from a single flat interface, the GH effect is much smaller than for beams reflected from periodically corrugated or multilayered structures, in which the shift could be of the same order of magnitude as the beam width \cite{tamirbertoni}. 
Apart from the case of linear isotropic dielectric media, nonspecular phenomena have been estimated and measured for other kind of homogeneous materials, such as chiral \cite{Bonomo82}, anisotropic uniaxial \cite{Bonomo57,Bonomo73,Bonomo75} or dielectric--magnetic media with negative refraction index \cite{berman,NPVLakhtakiaSLABS,NPVLakhtakia}. 
A unified linear algebra approach to study the generalized shifts for the polarization components of reflected light beams is presented in \cite{NJP-gottedennis}. 
For a self-consistent description of the GH shift and related beam-shift phenomena together with an overview of their most important extensions and generalizations, the reader is referred to \cite{BliokhAiello}. 

For total internal reflection from a boundary between two positive index media, the GH shift is typically the same order of magnitude as the wavelength and positive (i.e., the reflected beam is  shifted to the other side of the normal from the incident beam). 
On the contrary, for reflection from a metallic boundary and provided that the beam is p polarized, the GH shift may be negative (i.e., the reflected beam can be shifted towards the same side of the normal as the incident beam). 
As shown in Ref. \cite{Lai}, this negative shift is due to the fact that the p polarized evanescent wave in the metal is {\em backward}, i.e., it has an energy flow opposite to the direction of the phase velocity along the interface. 
For total internal reflection at the boundary between a positive index medium and a negative index medium  
\cite{berman,NPVLakhtakiaSLABS,NPVLakhtakia}, the lateral shift is also negative because the evanescent wave is again backward \cite{ishimaru1}. 

Although the beam shift at a single flat interface can be either positive or negative depending on the type of the surface waves (forward or backward) excited by the incoming beam, the same is not true for beam reflection from other structures. For example, it is well known (see \cite{physicalorigin} and references therein) that the excitation of a forward surface wave in a prism-waveguide coupling system can result in either a positive or a negative GH 
shift. However, the erroneous belief that the excitation of a backward surface wave always results in a negative GH shift sometimes appears in the literature, particularly when multilayers with negative index materials are involved \cite{Shadrivov1,Chen,mentira}. 

The purpose of this paper is twofold: 
(a) to study the lateral displacement of spatially limited beams reflected from Attenuated Total Reflection (ATR) structures under conditions of resonant excitation of 
backward surface plasmon polaritons (SPPs) and 
(b) 
to apply the phenomenological model presented in \cite{zeller_incidente} 
to clearly evidence that the excitation of backward surface waves in multilayers with negative index metamaterials can lead to both negative and positive (and not exclusively negative) GH shifts, 
depending on whether the thickness of the coupling layer is higher or lower than a critical value representing a real zero of the reflection coefficient. 

The plan of this paper is as follows. In Section \ref{teoria} we present some technical background and two alternative methods that will be used to numerically calculate the GH lateral shift, namely the stationary phase approximation and the rigorous calculation of the field distribution of the reflected beam. 
In Section \ref{zero-pole} we discuss the GH shift near conditions of resonant coupling between the incident field and backward surface waves in terms of the behavior of the complex poles and zeroes of the reflection coefficient \cite{zeller_incidente}. 
We give examples that show the existence of a critical thickness of the coupling layer: above this critical value the lateral shift is positive, whereas below this critical value the lateral shift is negative. 
In Section \ref{resultados} we present rigorous calculations of the transverse distribution of the reflected electric field near the resonant excitation of the backward surface plasmon polariton. We investigate the influence of the angle of incidence and the thickness of the ATR device for the optimal condition of coupling between the incident radiation and backward SPPs. Large  (negative or positive) GH shifts are obtained, accompanied by a splitting of the reflected beam. 
In Section \ref{conclusiones} we summarize and discuss the obtained results.

\section{GH shift calculation} \label{teoria} 

\begin{figure}[htbp!]
\centering
\resizebox{0.50\textwidth}{!}
{\includegraphics{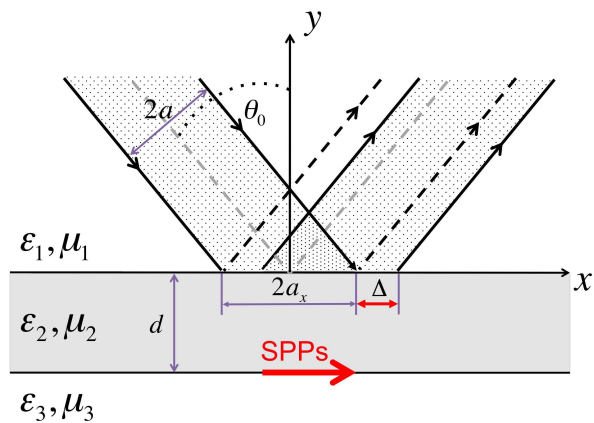}}
\caption{Schematic diagram of the system. It shows the lateral shift $\Delta$ of a bounded beam at an interface. The dashed lines show the boundaries of the reflected beam as predicted by geometrical optics theory.}
\label{sistema}
\end{figure}
Since the spatial periodicity associated with surface plasmon polaritons is less than the spatial periodicity which could be induced by an incident photon on the surface, surface plasmon polaritons cannot be resonantly excited by plane waves. This difficulty can be overcome by using ATR techniques, which require the introduction of a second surface, usually the base of an isosceles prism. In the Kretschmann configuration, the prism is positioned against the plasmonic medium (metal o metamaterial), while in the Otto configuration \cite{ottoNC} the prism is positioned very close to the surface of the plasmonic medium. 
Figure \ref{sistema} shows an Otto--ATR structure illuminated by a spatially limited gaussian beam (width $2a$).   
The beam is incident upon the base of the prism at $y=0$, the center of the beam waist is located at $x=0$ and the angle of incidence of the beam axis is $\theta_{0}$.  
Medium 3 is a magnetic metamaterial with complex values of electric permittivity $\varepsilon_{3}$ and magnetic permeability $\mu_{3}$. These constitutive parameters satisfy the condition for power flow and phase velocity in opposite directions \cite{MOTL41_315AnewConditionNPV}.  
Medium 1 (the prism) and medium 2 (the coupling layer, thickness $d$) are nonmagnetic ($\mu_{1}=\mu_{2}=1$) and have real and positive electric permittivities ($\varepsilon_{1}$ and $\varepsilon_{2}$ respectively). 
When the values of the relative constitutive parameters $\varepsilon_{3}/\varepsilon_{2}$ and $\mu_{3}/\mu_{2}$ are properly chosen \cite{darmanyanOC}, the interface 2--3 between the coupling layer and the metamaterial can support SPPs with time-averaged Poynting vector directed opposite to the phase velocity (backward SPPs). The examples of this paper are for the case $\Re \;(\varepsilon_{3}/\varepsilon_{2}) < -1$, that corresponds to $s$-polarized backward SPPs. 

When s-polarized incident radiation reaches the base of the prism with an angle $\theta_{0}$ greater than the critical angle of total reflection, the evanescent field can resonantly couple with the $s$-polarized backward SPP of the interface 2--3.
We assume that the electric field of the incident beam is given by
\begin{equation}
\textbf E_{i}= E_{i}(x,y=0)\,\hat z =\exp{[-(\frac{x}{a_{x}})^2+i\,\alpha_{0}\,x]}\,\hat z \,, \label{campoincidente1}
\end{equation}
with $\alpha_{0}=\frac{\omega}{c}\sqrt{\varepsilon_{1}\mu_{1}}\sin \theta_{0}$,  
$c$ the speed of light in vacuum, 
and $a_{x}=a\,\sec\,(\theta_{0})$. 
An $\exp{(-i\omega t)}$ time dependence is implicit, with $\omega$ the angular frequency, $t$ the time and $i =\sqrt{-1}$. 
Using a Fourier integral representation, the incident field can be written as 
\begin{equation}
E_{i}(x,y=0)=\int_{-\infty}^{\infty} A(\alpha)\exp{(i\,\alpha x)}\,d\alpha,
\label{campoincidente}
\end{equation}
where 
\begin{equation}
A(\alpha)=\frac{a_{x}}{2\sqrt{\pi}}\exp{[-(\alpha-\alpha_{0})^{2}\,(\frac{a_{x}}{2})^{2}]}\,,
\label{espectro}
\end{equation}
is the angular spectral distribution of the incident beam. 
The electric field of the reflected beam 
at the interface $y=0$ is given by 
\begin{equation}
E_{r}(x,y=0)
=\int_{-\infty}^{\infty}R(\alpha)\,A(\alpha)\,\exp{(i\,\alpha\,x)}\,d\alpha \,
\label{ref}
\end{equation}
where $R(\alpha)=|R(\alpha)| \exp{i\varphi(\alpha)}$ is the complex reflection coefficient of the multilayer as a function of the spectral variable $\alpha=\frac{\omega}{c}\sqrt{\varepsilon_{1}\mu_{1}}\sin \theta$. 
The beam shift can be obtained by finding the location where $|E_{r}(x,y=0)|$ is maximal. 
Using the stationary-phase approximation and assuming a linear variation for $\varphi(\alpha)$ and that the beam experiences total
internal reflection, this location is given by equation (\ref{stationary}). However, if $\varphi$ is not a linear function of $\alpha$ or if $|R(\alpha)| \neq 1$ along the spectral width of the incident beam, the reflected beam may suffer severe distortions and the stationary-phase result given by equation (\ref{stationary}) is usually unsuitable to describe the lateral shift correctly \cite{Tamir1973}.
In such cases, it is preferable to define the GH shift using the normalized first moment of the electric field of the reflected beam \cite{Hugonin} 
\begin{equation}
\Delta^{(1)}=\frac{\int_{-\infty}^{\infty}x|E_{r}(x,y=0)|^{2}dx}{\int_{-\infty}^{\infty}|E_{r}(x,y=0)|^{2}dx}\,. 
\label{corrimiento}
\end{equation}
In this definition, the shift is expressed as an average of the $x$-position of the reflected beam on the surface of interest and no approximation is done over the amplitude and phase of the reflected plane wave components of the beam. 
When the spectral width of the incident beam is small, it is possible to derive equation (\ref{stationary}) as a first order  approximation of equation (\ref{corrimiento}) \cite{artmanDEmomento}.

\section{Zero-pole model for the GH shift} \label{zero-pole}

To numerically illustrate 
the lateral displacement of spatially limited beams under conditions of resonant excitation of backward SPPs, we choose an Otto-ATR structure with $\varepsilon_{1}=2.25$, $\mu_{1}=1$, $\varepsilon_{2}=1$, $\mu_{2}=1$, $\varepsilon_{3}=-1.6+0.001\,i$ and $\mu_{3}=-0.7+0.001\,i$. For these constitutive parameters, the interface 2--3 between the coupling layer and the metamaterial supports $s$-polarized SPPs with time-averaged Poynting vector directed opposite to the phase velocity  \cite{darmanyanOC}. 
\begin{figure}[htbp!]
\centering
\resizebox{0.50\textwidth}{!}
{\includegraphics{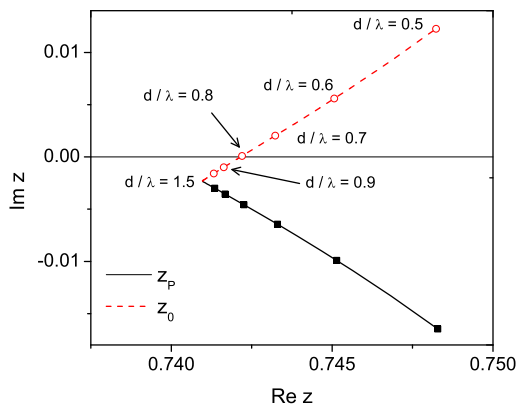}}
\caption{(Color online) Trajectories of the zero ($z_{0}$) and the pole ($z_{p}$) in the complex plane as functions of $d/\lambda$ for an Otto - ATR structure with $\varepsilon_{1}=2.25$, $\mu_{1}=1$, $\varepsilon_{2}=1$, $\mu_{2}=1$, $\varepsilon_{3}=-1.6+0.001\,i$ and $\mu_{3}=-0.7+0.001\,i$. For these constitutive parameters, the interface 2--3 supports $s$-polarized, backward SPPs.}
\label{ceroypolo}
\end{figure}

As shown by equation (\ref{ref}) the distortions of the reflected beam are ruled by the behavior of $R(\alpha)=|R(\alpha)| \exp{i\varphi(\alpha)}$. The exact form of this reflection coefficient is given for instance by equation (1) in \cite{zeller_incidente}. 
However, to clearly evidence the facts that i) the excitation of backward surface waves can lead to both negative and positive (and not exclusively negative) GH shifts and that ii) the sign of the GH shift depends on wheater the value of the coupling layer thickness is higher or lower than the critical value characteristic of the ATR structure, first we choose to represent this quantity within the frame of a phenomenological model based on the properties of the complex poles and zeroes of the reflection coefficient. 
According to this model, used previously for the case of Kretschmann-ATR structures with metamaterials \cite{zeller_incidente}, the reflection coefficient $R$ can be written as 
\begin{equation}
R (z,d/\lambda)= \zeta(z,d/\lambda)\,\frac{z-z_{0}(d/\lambda)}{z-z_{p}(d/\lambda)},
\label{rmodelo}
\end{equation}
where $z$ is the analytic continuation of $\sin \theta$ to the complex plane, $z_{0}$ and $z_{p}$ denote respectively the complex zero and the complex pole of $R$, $\lambda$ is the wavelength in vacuum and $\zeta(z,d/\lambda)$ is a complex regular function near $z_{0}$ and $z_{p}$ that does not change significantly near $z_{p}$.

\begin{figure}[htbp!]
\centering
\resizebox{0.50\textwidth}{!}
{\includegraphics{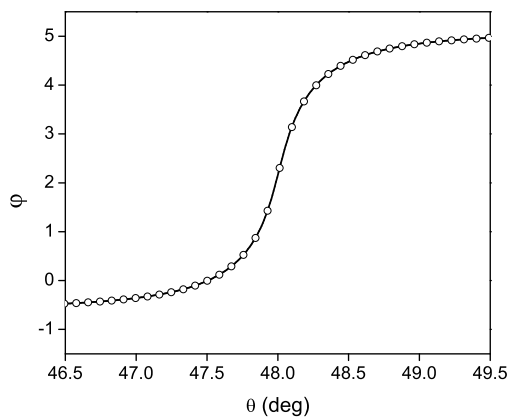}}
\caption{Phase $\varphi$ of the reflection coefficient for an Otto-ATR structure with $\varepsilon_{1}=2.25$, $\mu_{1}=1$, $\varepsilon_{2}=1$, $\mu_{2}=1$, $\varepsilon_{3}=-1.6+0.001\,i$, $\mu_{3}=-0.7+0.001\,i$ and $d/\lambda=0.7$.}
\label{d07}
\end{figure}
As explained in \cite{zeller_incidente} (see figures 2 and 3 therein), one of the main characteristics of using the zero-pole model to represent the optical response near the excitation of SPPs is that it provides a direct visualization of the behavior of the phase curves $\varphi(\alpha)$ in terms of the spectral variable $\alpha$. 
This characteristic makes the zero-pole model particularly suited for the study of the lateral GH shift, since the quantity that fixes the sign of the GH shift in the stationary phase approximation equation (\ref{stationary}) is precisely the derivative of the phase.  

The interested reader is referred to \cite{zeller_incidente} for further details, here we limit ourselves to mentioning the key result regarding the sign of the GH shift, namely that phase curves $\varphi(\alpha)$ have a very different behavior depending on the location in the complex plane of both the zero and the pole: when $\Im(z_{0})\,\Im(z_{p})<0$, the phase curve $\varphi(\alpha)$ is a monotonically increasing function, whereas when $\Im(z_{0})\,\Im(z_{p})>0$, the phase curve is a monotonically decreasing function, exhibiting first a maximum and then a minimum. In terms of the sign of the GH shift, negative GH shifts should be expected in the first case, whereas positive GH shifts should be expected in the second case. 
\begin{figure}[htbp!]
\centering
\resizebox{0.50\textwidth}{!}
{\includegraphics{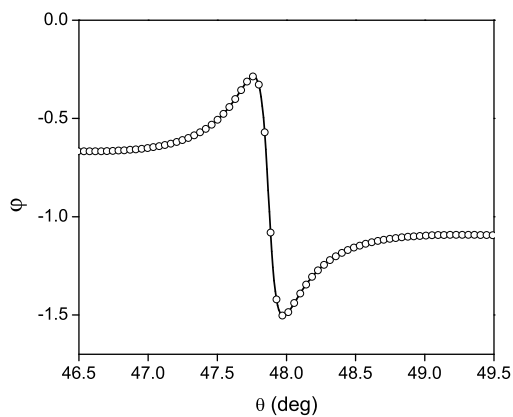}}
\caption{
%(Color online) 
Phase $\varphi$ of the reflection coefficient for an Otto-ATR structure with $\varepsilon_{1}=2.25$, $\mu_{1}=1$, $\varepsilon_{2}=1$, $\mu_{2}=1$, $\varepsilon_{3}=-1.6+0.001\,i$, $\mu_{3}=-0.7+0.001\,i$ and $d/\lambda=0.9$.}
\label{d09}
\end{figure}

Figure \ref{ceroypolo}, where we plot the parametric trajectories of $z_{0}(d/\lambda)$ and $z_{p}(d/\lambda)$ calculated for the values of the constitutive parameters used in this example, clearly shows that both cases ($\Im(z_{0})\,\Im(z_{p})<0$, negative GH shift, or $\Im(z_{0})\,\Im(z_{p})>0$, positive GH shift) can occur near the excitation of a backward SPP. 
The pole trajectory cannot cross the real axis, (if it did, infinite reflectance would result for a real angle of incidence), but the trajectory of the zero of the reflection coefficient is not limited and in this example it crosses the real axis for a critical value of $d/\lambda \approx 0.8$. 
The value of $z_{0}$ ($\approx 0.74$) at this critical thickness of the coupling layer corresponds to an angle of incidence $\theta_{0}=47.8^{\circ}$ for which a plane wave is totally absorbed by the ATR structure. 
We conclude that, for values of $d/\lambda$ lower than this critical value, the zero-pole model predicts that the function $\varphi(\alpha)$ is monotonically increasing (as the curve shown in figure \ref{d07} calculated for $d/\lambda=0.7$), and according to the stationary phase approximation, negative GH shifts are to be expected. On the other hand, for values of $d/\lambda$ above this critical value, the zero-pole model predicts that the function $\varphi(\alpha)$ is monotonically decreasing (as the curve shown in figure \ref{d09}, calculated for $d/\lambda=0.9$), and according to the stationary phase approximation the GH shift will be positive. 

\begin{figure}[htbp!]
\centering
\resizebox{0.50\textwidth}{!}
{\includegraphics{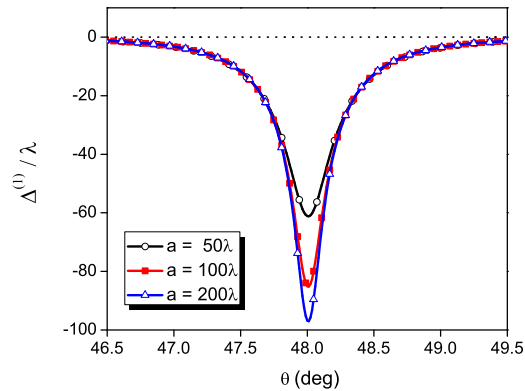}}
\caption{
(Color online) 
Lateral shift as a function of the angle of incidence for $d/\lambda=0.7$ and different widths of the gaussian incident wave.}
\label{fig1}
\end{figure}

\section{Transverse structure of the reflected beam} \label{resultados}

To verify the prediction from the zero-pole model and the stationary phase approximation that the sign of the GH shift near the excitation of a backward surface wave depends critically on the thickness of the coupling layer, we compare the results of the previous section with those obtained from the rigorous calculation of the spatial structure of the reflected beam (equation (\ref{ref})). The integrals in equation (\ref{ref}) and (\ref{corrimiento}) have been numerically calculated. 

\begin{figure}[htbp!]
\centering
\resizebox{0.50\textwidth}{!}
{\includegraphics{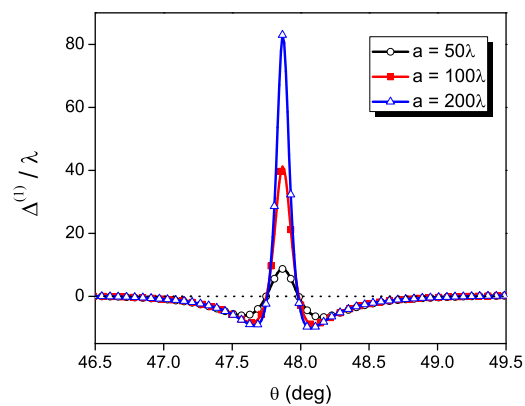}}
\caption{(Color online) 
Lateral shift as a function of the angle of incidence for $d/\lambda=0.9$ and different widths of the gaussian incident wave.}
\label{fig2}
\end{figure}
Figure \ref{fig1} (for $d/\lambda=0.7$, a value lower than the critical value) and \ref{fig2} (for $d/\lambda=0.9$, a value  higher than the critical value) show curves of $\Delta^{(1)}$, the normalized first moment of the reflected electric field given by equation (\ref{corrimiento}), as functions of the angle of incidence $\theta$ and for different values of the beam width $a$. 
In all the cases, the sign of $\Delta^{(1)}$ is in complete agreement with the 
zero-pole model and the stationary phase approximation. 
For both values of $d/\lambda$, the larger the width of the incident beam, the greater the peak value of $|\Delta^{(1)}|$. 
Due to the shift of the SPP resonance with the coupling layer thickness, peak values occur for different angles of incidence 
($\theta\approx48^{\circ}$ for $d/\lambda=0.7$ and $\theta\approx47.9^{\circ}$ for $d/\lambda=0.9$).  
Note that large peak values for $|\Delta^{(1)}|$ are obtained in both cases. When $d/\lambda=0.7$, the peak values of $\Delta^{(1)}/\lambda$ are $-61.18$, $-89.19$ and $-97.02$ for $a/\lambda=50, 100$ and $200$ respectively, whereas when $d/\lambda=0.9$, the peak values of $\Delta^{(1)}/\lambda$ are $8.67$,  $41.18$ and $80.53$ for $a/\lambda=50$, $100$ and $200$ respectively.

\begin{figure}[htbp!]
\centering
\resizebox{0.50\textwidth}{!}
{\includegraphics{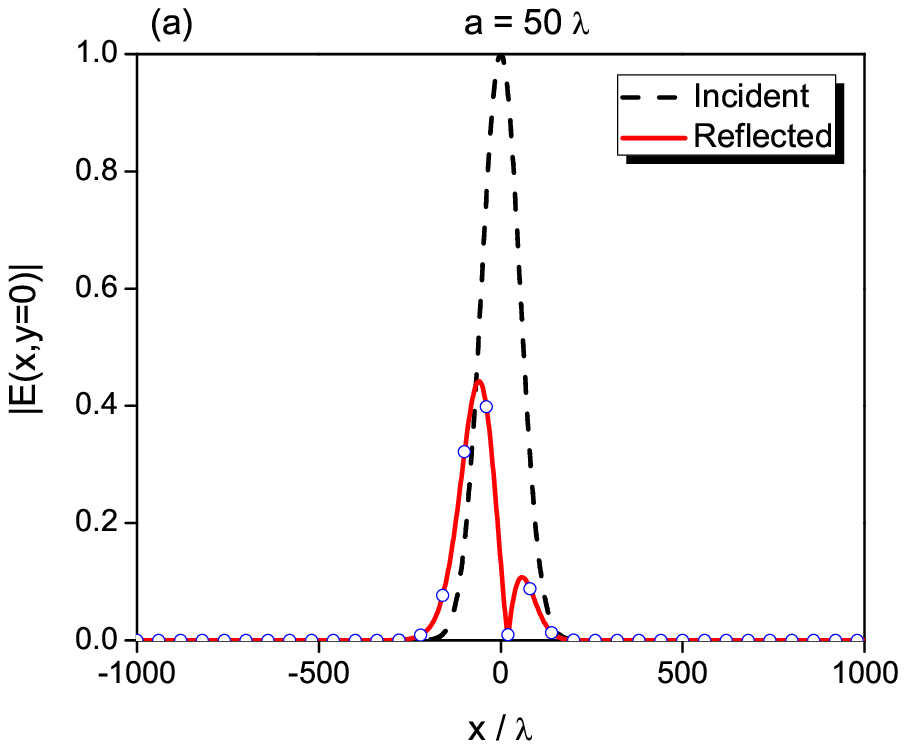}}
\resizebox{0.50\textwidth}{!}
{\includegraphics{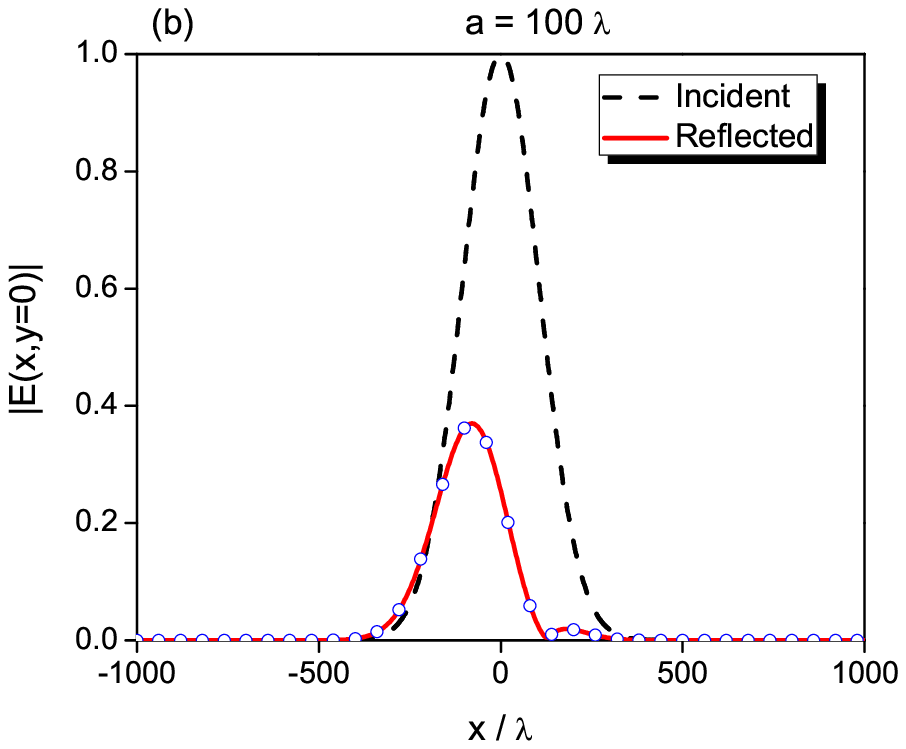}}
\resizebox{0.50\textwidth}{!}
{\includegraphics{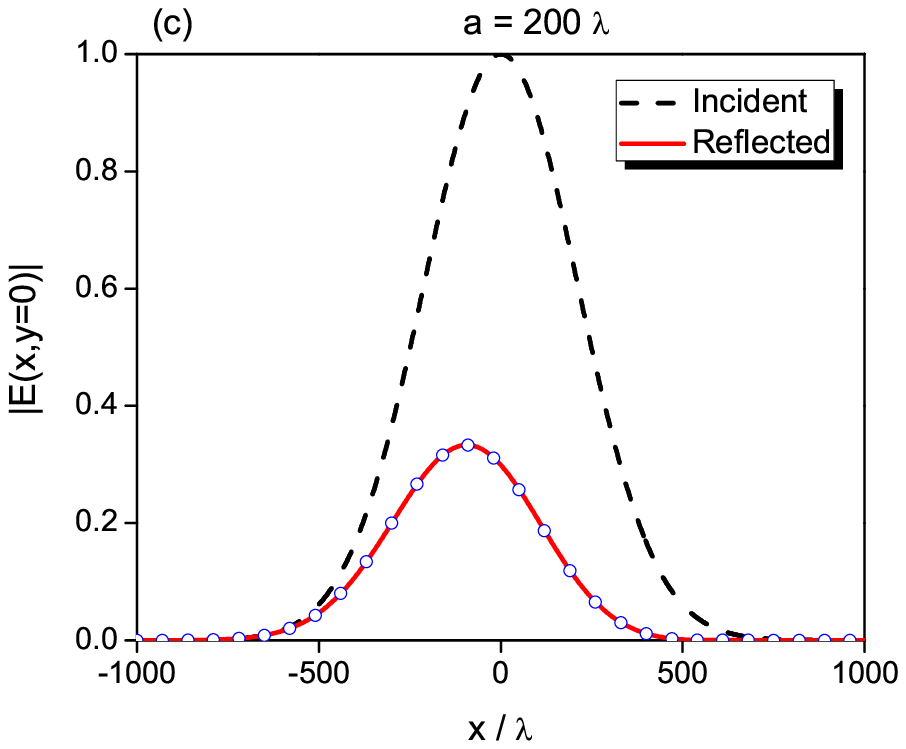}}
\caption{(Color online) 
Profiles of $|E_{r}(x,0)|$ calculated with equation (\ref{ref})(solid line), $|E_{r}(x,0)|$ calculated with equation 
(\ref{E-r6}) (circles) and the Gaussian incident field (dashed line) for $d/\lambda=0.7$, $\theta=48.01^{\circ}$ and values of $a/\lambda$ considered in figure \ref{fig1}.}
\label{perfiles07}
\end{figure}

\begin{figure}[htbp!]
\centering
\resizebox{0.50\textwidth}{!}
{\includegraphics{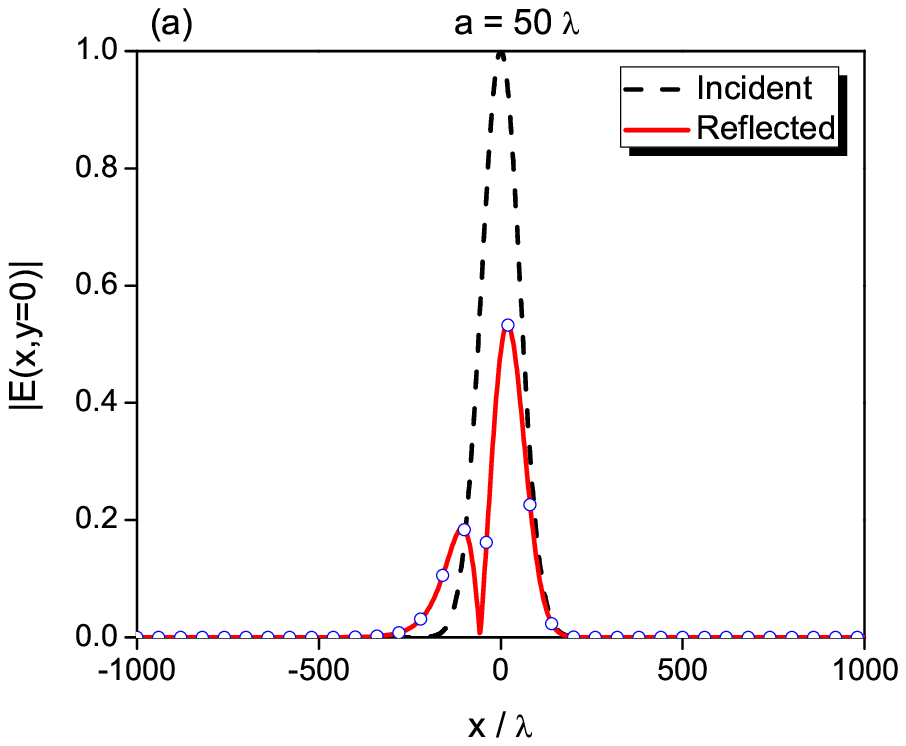}}
\resizebox{0.50\textwidth}{!}
{\includegraphics{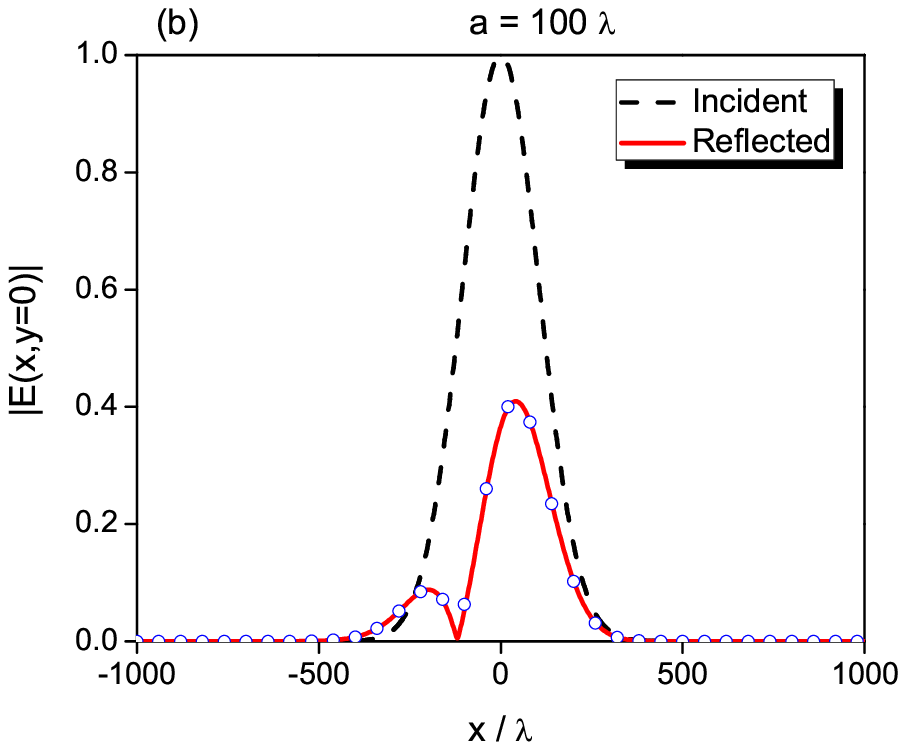}}
\resizebox{0.50\textwidth}{!}
{\includegraphics{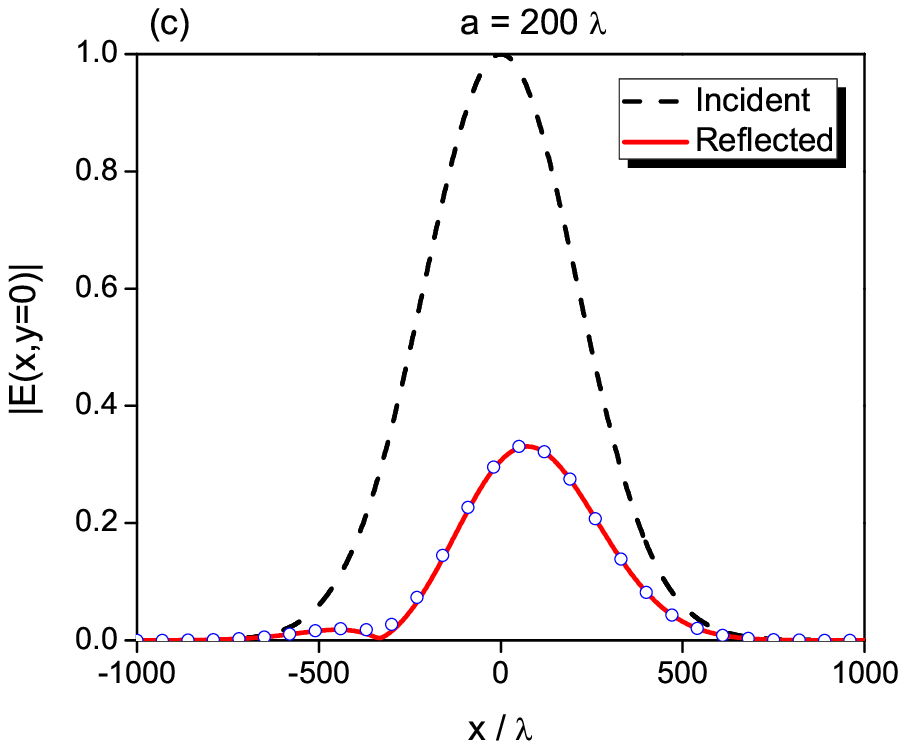}}
\caption{(Color online) 
Profiles of $|E_{r}(x,0)|$ calculated with equation (\ref{ref})(solid line), $|E_{r}(x,0)|$ calculated with equation 
(\ref{E-r6}) (circles) and the Gaussian incident field (dashed line) for $d/\lambda=0.9$, $\theta=47.86^{0}$  and values of $a/\lambda$ considered in figure \ref{fig2}.}
\label{perfiles09}
\end{figure}

Under conditions as those considered in these examples, such as total absorptions and resonant excitation of SPPs, the transverse structure of the reflected beam can have a complicated form. Particularly, the spatial distribution in the beam can be asymmetric, so that the value of the shift defined by equation (\ref{stationary}) may not coincide with the value obtained from the first moment of the spatial structure of the reflected beam. 
% aca lo de Tamir para el 2do referi
In these cases, and following a procedure similar to that presented in a paper by Tamir and Bertoni \cite{tamirbertoni}, it is possible to obtain a description which i) accounts for the deformation of the beam, ii) agrees quantitatively with the numerical calculations, and iii) takes advantage of the phenomenological model described by equation (\ref{rmodelo}). 
The advantage of using the zero-pole model to represent the reflection coefficient is that the integrals in 
equation (\ref{ref}) can be evaluated explicitly in terms of the complementary error function of complex argument \cite{tamirbertoni}. In this description, the electric field of the reflected beam at the interface $y = 0$ results 
\begin{equation}
E_r (x,0)=\zeta\;E_{i}(x,0)\left[1- i\;\pi^{3/2} n_1 (z_p-z_0) \frac{a_x}{\lambda}\;\exp(\gamma^2)\;\mathrm{erfc}(-\gamma)\right]\,,
\label{E-r6}
\end{equation}
where $n_1=\sqrt{\varepsilon_{1}\mu_{1}}$ and $\mathrm{erfc}$ is the complementary error function evaluated at the complex argument $\gamma$ 
\begin{equation}
\gamma=-\frac{x}{a_x}+i\frac{a_x}{\lambda}\pi n_1(z-z_p) \,.
\label{gama}
\end{equation}
Thus, the reflected beam is obtained as a product of two terms: one with a Gaussian form, which is multiplied by another function so that the reflected field is no longer Gaussian. 
To gain a deeper understanding of the distortions of the reflected beam, we have calculated the transverse distribution of the reflected electric field as a function of the $x$-coordinate. As shown in Figures \ref{perfiles07} and \ref{perfiles09}, the results obtained using equation (\ref{ref}) (solid lines) are in very good agreement with those obtained using equation (\ref{E-r6}) (circles). 

%las figuras tienen agregado el calculo de Tamir
Figure \ref{perfiles07} shows $|E_{r}(x,0)|$ (equation (\ref{ref})) for $d/\lambda=0.7$, $\theta=48.01^{\circ}$ (the angle of incidence for which the peaks in $\Delta^{(1)}$ occur, see figure \ref{fig1}) and for different values of the beam width $a$. The spatial distribution of the incident electric field (dashed line) is given as a reference. 
In these examples we observe that the spatial distributions 
are always shifted toward negative values of $x$, in total agreement with the 
values of $\Delta$ and $\Delta^{(1)}$ obtained for a structure with the coupling layer thickness lower than the critical value.  However, the profiles of the reflected fields are distorted with respect to the Gaussian profile of the incident beam, and the more severe distortion is obtained for the smallest beam width. The profiles exhibit a splitting of the reflected beam. As explained in \cite{Tamir}, the major peak is produced by the spectral components of the incident beam which are coupled to the excited SPP, while the minor peak is produced by the spectral components of the incident beam which are not coupled with SPPs. 

For a structure with the coupling layer thickness higher than the critical value, the spatial distribution of reflected electric field is expected to be shifted toward positive values of $x$. The rigorous calculation of $|E_{r}(x,y=0)|$ confirms this expectation, as shown in figure \ref{perfiles09}, analogous to figure \ref{perfiles07}, but for $d/\lambda=0.9$ and 
$\theta=47.86^{\circ}$ (the angle of incidence for which the larger values for $\Delta^{(1)}$ occur, see figure \ref{fig2}). 
Except for the sign of the lateral shift, the features exhibited by the profiles in figure \ref{perfiles09} are completely analogous to those in figure \ref{perfiles07}. As in the previous example, a double peak structure in the reflected field  profile is present and the more severe distortions occur for the smallest beam width.

\section{Conclusions} \label{conclusiones}

We have presented an exhaustive study of the GH lateral shift in Otto-ATR systems with negative index metamaterials near conditions of resonant coupling between the incident field and backward surface waves. 
Although the GH beam shift at a {\em single flat interface} can be either positive or negative depending on the type of surface waves (forward or backward) excited by the incoming beam, the erroneous belief that the excitation of a backward surface wave always results in a negative GH shift frequently appears in the literature, particularly when multilayers with negative index materials are involved. To clearly evidence that the excitation of backward surface waves in multilayers with negative index metamaterials can lead to both negative and positive (and not exclusively negative) GH shifts, we have adapted to the Otto configuration the zero-pole model presented in \cite{zeller_incidente} for the Kretschmann configuration. This model  predicts the existence of a critical thickness of the coupling layer: above this critical value the lateral shift is positive whereas below this critical value the lateral shift is negative. We have presented rigorous calculations of the transverse  distribution of the reflected electric field near the resonant excitation of the backward surface plasmon polariton. These calculations show the existence of large (negative or positive) Goos-H\"anchen shifts accompanied by a splitting of the reflected beam and  confirm the predictions of the zero-pole model.

\section*{Acknowledgment}
The authors acknowledge the financial support of Consejo Nacional de Investigaciones Cient\'{\i}ficas y T\'ecnicas, (CONICET, PIP 451)
and Universidad de Buenos Aires (project UBA 20020130100718BA).

\end{document}